\documentclass[12pt]{article}

\usepackage{amsmath,amsfonts}
\usepackage{graphicx}
\usepackage{epsfig}
\newcount\rys

\begin{document}

\title{Topological quantum numbers and curvature ---
examples and applications}

\vspace{0.6cm}


\author{
Jerzy Szcz\c{e}sny \\
Institute of Physics. Pedagogical University of Cracow,\\
Podchor\c{a}\.zych 2, 30 - 084 Cracow, Poland\\
Marek Biesiada \\
Department of Astrophysics and Cosmology, \\
Institute of Physics, University of Silesia,\\
 Uniwersytecka 4,
40-007 Katowice, Poland \\
Marek Szyd{\l}owski \\
Astronomical Observatory, Jagiellonian University,\\
Orla 171, 30-244 Krak{\'o}w, Poland; \\
Marc Kac Complex Systems Research Centre, Jagiellonian University,\\
Reymonta 4, 30-059 Krak{\'o}w, Poland}

\date{}
\maketitle
\begin{abstract}
 Using the idea of the degree of a smooth mapping between two manifolds of the same dimension we present here the topological (homotopical)
 classification of the mappings between spheres of the same dimension, vector fields, monopole and instanton solutions.
 Starting with a review of the elements
of Riemannian geometry we also present an original elementary
proof of the Gauss-Bonnet theorem and the Poincar\'{e}-Hopf theorem.

\end{abstract}

\section{Introduction}
From the beginning of the first half of the 1970s we can observe
the growing importance of the topological (global) methods of
analysing the structures of solutions of nonlinear equations of
mathematical physics \cite{{ref_Manton}}.  

In this article we discuss first the very important idea of the degree of a smooth mapping between smooth manifolds,
a homotopical invariant which is very useful for the above mentioned methods.

We describe the fundamental properties of degree of a smooth
mapping between the manifolds of the same dimension. In relatively
simple way we introduce the integral formula for the mapping
degree and in a systematic way we present explicit formulas for
the degree of the mapping between the $n$-dimensional spheres.

We then determine, in a new way, the explicit formulas for the index of vector field in relation to hypersurfaces and to point.
The novelty of our approach lies in an effective use of differential forms.

Using the idea of an index of a vector field we then present a
formula for the topological quantum number characterizing monopole
configurations in the Yang-Mills-Higgs theory with gauge
group $SU(2)$. Using the idea of an index of a vector field we
derive the formula for a ``topological quantum number'' (Chern
number) characterizing homotopically nonequivalent instanton
configurations of Yang-Mills theory with gauge group $SU(2)$.
We show also the bijectivity between the elements of the $SU(2)$
gauge group and unit vector fields on $R^4$

 After introducing the elements of Riemannian geometry we present an original, elementary proofs of Gauss-Bonnet and Poincar\'{e}-Hopf
 theorems for compact, closed, oriented, two-dimensional manifolds. The traditional proof of the Gauss-Bonnet theorem uses classical geometry
 and is much more complicated. Proving the Gauss-Bonnet
theorem we used effective language of the differential forms and
skilfully used the notion of the form of connection. The new
ingredient in the proof of the Poincar\'{e}-Hopf theorem is the
expression of an index of the vector field at a point by suitable
scalar products (formula (23) below). This line of proof (after a
generalization to higher dimensions) creates a possibility of
formulation a sufficient condition for the existence of a
pseudo-Riemannian metric on compact manifolds without a boundary.
Such a condition would be very useful in modern cosmology. Finally
we discuss the Poincar\'{e}-Hopf theorem for any even
dimensional compact orientable manifold without boundary.

Instantons are very interesting both from the physical and
mathematical point of view. They are topologicaly nontrivial
solutions of Yang-Mills equations which globally minimized in
chosen topological class of a functional of euclidean action. One
can use them to analyze tunneling processes occurring in different
systems, e.g. in Yang-Mills theory and in the system described
by nonrelativistic quantum mechanics
\cite{ref_Rajamaran} 
 The topological characterization of monopoles and instantons, presented in this paper, might be useful in many applications.
 To remind the importance
 of monopoles in physics, let us mention that the so called 'monopole problem' was a direct motivation for the inflationary models in cosmology
 \cite{ref_Linde}, 
 and that monopoles can catalyse the decay of a proton \cite{ref_Rubakov}. 

In addition, this work presents an elementary topological classification of two-dimensional surfaces via a heuristic proof of the Gauss-Bonnet theorem.
Such classifications are commonly used in string theory.

\section{The basic definitions}

\subsection{Critical points of the mapping}
We will consider a smooth mapping $f \colon M \to N$ between the smooth manifolds $M$ and $N$. By $C_f$ we denote the collection of points $p \in M$ such
that differential (tangent mapping) $d_p f \colon T_p M \to T_{f(p)} N$ has a smaller rank than the dimension of the manifold $N$:
\begin{displaymath}
C_f = \left\{ {p \in M \colon rk\left({d_p f} \right) < \dim N} \right\}
\end{displaymath}\\
Here the set $C_f$ is the set of the critical points of the $f$ mapping and the set $f(C_f)$ is the set of the critical values of this mapping.\\
\textit{Definition}\\
The set $A \subset R^n$ is called the set of the null measure in $R^n$ if one can cover it with a countable family of the $n$-dimensional cubes with
the arbitrary, small total volume of that cover.
We can generalize the definition of the set of the null measure also for subsets of the smooth manifolds, namely the subset $A \subset M$, where $M$
is $n$-dimensional smooth manifold with the atlas $\left\{ {(U_i,\phi _i)} \right\}$.\\
We say that $A$ has a  null measure if for any coordinate mapping
$\phi _i :U_i \to R^n$ the image $\phi _i \left({U_i \cap A}
\right)$ has a null measure set in $R^n$. The fundamental Sard's
theorem says \cite{ref_Dubrovin}: Let $f \colon M \to N$ be the smooth
mapping of the smooth manifolds $M$ and $N$. Then the set of the
critical value $f(C_f)$ of this mapping is the null measure
set in $N$.

A point $p \in M$ is called a regular point of the smooth mapping $f \colon M \to N$ if it is not the critical point, i.e. when $rk(d_p f) = \dim N$.

A point $q \in N$ is called a regular value of the mapping $f \colon M \to N$ if all points $p_a \in M$ belonging to the inverse image
$f^{ - 1} (\left\{ q \right\})$ of the point $q$ are the regular points of the mapping $f$. 
(When $f^{ - 1} \left({\left\{ q \right\}} \right) =$ $\emptyset$ then the point $q$ we call the regular value of the mappings $f$.)
 
\subsection{Homotopy}
The smooth (continuous) homotopy of the mapping $f \colon M \to N$ we call the smooth (continuous) mapping $F$ of the cylinder $M \times \left[ {0,1} \right]$ in
manifold $N$ i.e. $F \colon M \times \left[ {0,1} \right] \to N$ such that $F(p,0) = f(p)$ for any point $p \in M$.

About the mappings $f_t \colon M \to N$ where $f_t (p) = F(p,t)$ we say that they are homotopic with the ''initial'' mapping $f_0 = f$. For any mapping of
the cylinder $M \times [0,1]$ we say that it is the homotopy or the homotopy process between the mappings $f_0$ and $f_1$ where $f_1 (p) = F(p,1)$. We say
two mapping $f_0$ and $f_1$ are homotopic when there is the homotopy between them.

The set of points of the form $(p,0)$ we call the base of the cylinder while the set of points of the form $(p,1)$ we call its bottom.

Therefore two mappings $f_0 \colon M \to N$ and $f_1 \colon M \to N$ are homotopic if there exists a smooth (continuous) mapping of the cylinder over $M$ that extends
the map defined by $f_0$ at the bottom of the cylinder and by $f_1$ defined at the top.
The same way all mappings from $M$ to $N$ share homotopy classes, and each class consists of the mappings homotopic with one another. The set of homotopy
classes of the mappings from $M$ into $N$ we denote by $\left\{ {M,N} \right\}$.\\ \\
\textit{Definition}\\
Two manifolds $M$ and $N$ we call homotopicaly equivalent if there exist the smooth mappings $f \colon M \to N$ and $g \colon N \to M$ such that the composite
mappings $g \circ f colon M \to M$ and $f \circ g \colon N \to N$ are homotopic with the identity mappings. If the manifolds $M$ and $N$ are homotopicaly equivalent
then for any manifold $P$ there exists a bijection between their respective homotopy classes $\left\{ {P,M} \right\}$ and $\left\{ {P,N} \right\}$.

\subsection{Degree of mapping}
The degree of mapping is a quantity with which we are able to consider the homotopy classes among the manifolds: closed (i.e. compact and without
boundaries), oriented, $M$ and $N$ with the same dimension; $\dim M = \dim N$.

Lets consider the smooth mapping $f \colon M \to N$ and fix a point $q_0 \in N$. We will assume that the mapping $f \colon M \to N$ is the proper mapping in relation
to the point $q_0$ i.e. the inverse image $f^{ - 1} \left({\left\{ {q_0 } \right\}} \right)$ consists of the finite number of points
$f^{ - 1} \left({\left\{ {q_0 } \right\}} \right) = \left\{ {p_1,\ldots,p_R } \right\} \subset M$ and for $a = 1,\ldots,R$,
$\det [d_{p_a } f] \ne 0$.\\\\
\textit{Definition}\\
The degree of the smooth mapping $f \colon M \to N$ $(\deg f)$ of the connected manifolds, closed, oriented, possessing the same dimension, is
defined by an integer
\begin{equation}
\deg f = \sum\limits_{p_a \in f^{ - 1} \left({\left\{ {q_0 } \right\}} \right)} {{\mathop{\rm sgn}} [\det \left({d_{p_a } f} \right)} ]
\label{ref_eq01}
\end{equation}
($f$ is the proper mapping in the relation to the point $q_0$).

The degree of the mapping we call also the algebraic number of inverse image. We quote without the proof  the very important theorem of the
degree of mapping:\\\\
\textit{Theorem}\\
The mapping degree does not depend on the choice of the regular value $q_0$ and is invariant under a homotopy process i.e. does not change with
the homotopy of the given mapping---in other words---it is the characteristics of the element belonging to the set $\left\{ {M,N} \right\}$.

Heuristically, the homotopical invariance of the degree can be shown via a continuity argument. Namely, the degree, by definition, takes integer
values and is a continuous function of a deformation parameter t, hence it must be a constant function of this parameter. It is clear from the
integral formula for the degree that it is independent of the choice of a regular value of the mapping.

Moreover, the following theorem is true:
Two smooth mappings $f$ and $g$ : $M \to S^n$ of an $n$-dimensional closed, oriented manifold $M$ to the $n$-dimensional sphere $S^n$ are homotopic
if and only if (iff) when their degrees overlie \cite{ref_Dubrovin}.

\section{Integration of the form and degree of mapping}

We will show that the degree of the mapping can be specified by
integrating the proper differential forms over the manifolds. Let
us consider the behavior of the integral of the differential
$n$-form $\omega$ over a $n$-dimensional closed and oriented
manifold which respect to the mapping $f$ with the specified
degree. Let $M$ and N be the manifolds of the same dimension n,
closed and oriented and let $f \colon M \to N$ be the smooth mapping
having the specified degree denoted by $\deg f$ and let $\omega$
be the differential $n$-form on $M$. We have  the following
formula:
\begin{equation}
\int\limits_M {f^ * } \omega = \deg f\int\limits_N \omega,
\label{ref_eq02}
\end{equation}
where $f^ * \omega$ is the pull-back of the form $\omega$ on the manifolds $M$ by the mapping $f$.\\
Sketch of the proof:\\
Let us choose in $N$ the regular value $q_0$ of the mapping $f$. Lets $\hat V$ denote the neighborhood of the point $q_0$ which is created by the
points being the regular values of the mapping $f$.\\
Let the inverse image $f^{ - 1} \left({\left\{ {q_0 } \right\}} \right)$ consists of the regular points $p_1^0,\ldots,p_R^0$ of mapping $f$. The
neighborhoods of these points $\hat U_1,\ldots,\hat U_R$
is the inverse image of the set $\hat V$ : $f^{ - 1} (\hat V) = \hat U_1 \cup \hat U_2 \cup\ldots \cup \hat U_R$.

Every neighborhood $\hat U_a$ $a =1,2,\ldots,R$ consists of the regular points of mapping $f$ and the subsets $\hat U_a$ with a suitable choice
of the neighborhood are disjoint.\\
On manifolds $M$ and $N$ we always choose the coordinate mappings such that the disjoint coordinate neighborhoods $U_a$ $(a = 1, 2,\ldots,R)$
fulfill the conditions $\hat U_a \subset U_a$.\\\\
The mappings $\phi _a \colon U_a \to$ $R^n$ define in the manifold $M$ local coordinates $\phi _a (p) = \left({x_a^1,\ldots,x_a^n } \right)$ of point
$p \in U_a$. Likewise, on manifold $N$ we choose the coordinate neighborhood $V$ in such a way that $\hat V \subset V$, with  local coordinates
defined by the mapping $\psi \colon V \to R^n$ denoted by $\left({y^1,\ldots,y^n } \right) = \psi (q)$ for $q \in V$.
In local coordinates $\left({y^1,\ldots,y^n } \right)$ the differential $n$-form $\omega$ has the form
$\omega = g\left({y^1,\ldots,y^n } \right)dy^1 \wedge dy^2 \wedge\ldots \wedge dy^n$ where $g$ is a smooth real-valued function.

The pull-back form $f^ * \omega$ in neighborhood of point $p_a^0$ in local coordinates has the form:
\begin{displaymath}
f^ * \omega \left| {_{\hat U_a } } \right. = g \circ f\det \left[\frac{{\partial f^\mu }}{{\partial x_a^\nu }}(x_a)\right]
dx_a^1 \wedge\ldots \wedge dx_a^n.
\end{displaymath}
The restriction of a mapping $f$ to subset $\hat U_a$ is the diffeomorphism $f \colon \hat U_a \to \hat V$. Hence:
\begin{eqnarray*}
\int\limits_{\hat U_a } {f^ * } \omega & = & \int\limits_{\phi _a (\hat U_a)} {g(f(x_a)})\det \left[\frac{{\partial f^\mu }}
{{\partial x_a^v }}(x_a)\right]dx_a^1 \wedge\ldots \wedge dx_a^n \\
& = &{\mathop{\rm sgn}} \left[\det \left({ {\frac{\partial f^\mu }{\partial x_a^\nu }(p_a^0)}} \right)\right]
\int\limits_{\phi _a (\hat U_a)} {g\left({f(x_a)} \right)} \left| {\det \left[\frac{{\partial f^\mu }}{{\partial x_a^\nu }}(x_a)\right]}
\right|{dx_a^1\ldots dx_a^n}^n \\
& = & {\mathop{\rm sgn}} \left[\det \left({\frac{\partial f^\mu}{\partial x_a^\nu}(p_a^0)} \right)\right]\int\limits_{\psi (\hat V)}
{g(y^1 },\ldots ,y^n)dy^1\ldots d y^n \\
& = & {\mathop{\rm sgn}} \left[\det (d_{p_a^0 } f)\right]\int\limits_{\hat V} \omega,
\end{eqnarray*}
where we used the fact that on the set $\phi _a (U_a)$ the sign of the determinant $\det \left({\frac{{\partial f^\mu }}
{{\partial x_a^\nu }}(x_a)} \right)$ is constant and we made use of the theorem on the change of variables in a multiple integral.
Hence:
\begin{displaymath}
\int\limits_{f^{ - 1} (\hat V)} {f^ * } \omega = \sum\limits_{a = 1}^R {\int\limits_{\hat U_a } {f^ * } } \omega =
\left[ {\sum\limits_{a = 1}^R {{\mathop{\rm sgn}} \left({\det (d_{p_a } f)} \right)} } \right]\int\limits_{\hat V} \omega.
\end{displaymath}
In other words:
\begin{displaymath}
\int\limits_{f^{ - 1} (\hat V)} {f^ * } \omega = \deg f\int\limits_{\hat V} \omega
\end{displaymath}
If this form $\omega$ has the values different from zero only on the set $\hat V$, then $\int\limits_{\hat V} \omega = \int\limits_N \omega$
and $\int\limits_{f^{ - 1} (\hat V)} {f^ * } \omega = \int\limits_M {f^ * \omega }$ so for this form:
\begin{displaymath}
\int\limits_M {f^ * } \omega = \deg \int\limits_N \omega.
\end{displaymath}
If we have any $n$-form defined on manifold $N$ we will see that on set $C_f \subset M$ form $f^ * \omega$ is equal to zero
so $\int\limits_M {f^ * \omega } = \int\limits_{M\backslash C_f } {f^ * \omega }$.
According to Sard's theorem the set $f(C_f)$ is the set of the null measure in $N$ so
\begin{displaymath}
\int\limits_N {\omega = \int\limits_{N\backslash f(C{}_f)} \omega }.
\end{displaymath}
Let $\left\{ {\alpha _i } \right\}$ be a partition of unity subordinated to the covering $\left\{ {V_i } \right\}$ of the manifold
$N\backslash f(C_f)$. Then the form $\alpha _i \omega$ has the value of null outside the set $V_i$ and
$\int\limits_M {f^ * (\alpha _i } \omega) = \deg f\int\limits_M {\alpha _i } \omega$.
After summation of the last equation with respect to $i$ we get:
\begin{displaymath}
\int\limits_{M\backslash C_f } {f^ * \omega = \deg f\int\limits_{N\backslash f(C_f)} \omega },
\end{displaymath}
because: $\sum\limits_i {\alpha _i } = 1$, $f^ * \left({\alpha _i \omega } \right) = (\alpha _i \circ f)f^ * \omega$ and
$\sum\limits_i {\alpha _i } \left({f(p)} \right) = 1$.
This concludes the proof of the theorem.

Let us remark that if $N$ is the closed manifold and $\eta$ is the volume form on $N$ i.e. when $\int\limits_N \eta = volN$ and
$f \colon M \to N$ is the smooth mapping between the closed and oriented manifolds of the same dimension then:
$\int\limits_M {f^ * } \eta = \deg f\int\limits_N {\eta = \deg f(volN})$, hence:
\begin{equation}
\deg f = \frac{1}{{volN}}\int\limits_M {f^ * } \eta
\label{ref_eq03}
\end{equation}
This integral formula provides an effective method to determine the degree of the mapping.

\section{Some applications of integral formula of degree of mapping}

\subsection{Solid angle form and spheres mapping}

In domain $R^{n + 1} \backslash \left\{ 0 \right\}$ which is homotopicaly equivalent with the sphere $S^n$ lets consider the following $n$-form:
\begin{equation}
\Omega _n = \frac{1}{{vol(S^n)}}\frac{{\sum\limits_{\alpha = 1}^{n + 1} {(- 1)^{\alpha - 1} x^\alpha dx^1 \wedge\ldots \wedge
d\mathop {x^\alpha }\limits^ \vee \wedge\ldots \wedge dx^{n + 1} } }}{{\left[ {(x^1)^2 +\ldots + (x^{n + 1})^2 } \right]^{\frac{{n + 1}}{2}} }}
\label{ref_eq04}
\end{equation}
where symbol ''$\vee$'' put over an object means that this object is omitted, $vol(S^n)$ is the area of the $n$-dimensional sphere:
\begin{eqnarray*}
vol(S^n) = \left\{ {\begin{array}{*{20}l}
{\frac{{2\pi ^{r + 1} }}{{r!}} - for - n = 2r + 1} \\\\
{\frac{{2^{r + 1} \pi ^r }}{{\left({2r - 1} \right)!!}} - for - n = 2r} \\
\end{array}} \right.
\end{eqnarray*}
It is easy to see that the form $\Omega _n$ is the closed form i.e. $d\Omega _n = 0$ but it is not the exact form because
$\int\limits_{S^n } {\Omega _n } = 1$ where $S^n$ is the $n$-dimensional sphere given by the equation $(x^1)^2 +\ldots + (x^{n + 1})^2 = 1$.
The form $\Omega _n$ is distinguished in the domain $R^{n + 1} \backslash \left\{ 0 \right\}$ in the sense that any closed $n$-form defined in
this domain has the form $Q\Omega _n + d\alpha$, where $Q$ is the real number and $\alpha$ is some $(n-1)$ differential form.

The geometrical meaning of the form $\Omega _n$ is as follows: If
some $n$-dimensional hypersurface $M^n$ in $R^{n + 1}$ is given by
the immersion $\phi \colon R^n$ $\supset U \to M^n$ i.e. by the
mapping $\underbrace {\left({t^1,t^2,\ldots,t^n } \right)}_{ \in
U} = t \mapsto \left({x^1 (t),\ldots x^{n + 1} (t)} \right) \in
M^n$, then the integral $\int\limits_U {\phi ^ * \Omega _n }$ says
what fraction of the total solid angle is covered by the
hypersurface $M^n$ when we look at it from the point $0$. An
example: for $n = 2$ we have:
\begin{eqnarray*}
\Omega _2 & = &\frac{1}{{4\pi }}\frac{{xdy \wedge dz + ydz \wedge dx + zdx \wedge dy}}{{\left| {\vec r} \right|^3 }}, \\
\phi ^ * \Omega _2 (t^1,t^2) & = &\frac{1}{{4\pi }}\frac{{\vec r(t) \cdot \vec n(t)}}{{\left| {\vec r(t)} \right|^3 }}\left|
{\vec E_1 \times \vec E_2 } \right|dt^1 \wedge dt^2,
\end{eqnarray*}
where $\vec E_a (t) = \frac{{\partial \vec r}}{{\partial t^a }}(t^1,t^2)$ $(a = 1,2)$ are vectors tangent to surface in $\vec r(t)$ point
when $\vec n(t) = \frac{{\vec E_1 (t)\times \vec E_2 (t)}}{{\left| {\vec E_1 \times \vec E_2 } \right|}}$ is the unit normal vector to surface
in point $\vec r(t)$.

Thus the form $\Omega _n$ can be called the normalized solid angle form.

The circle $S^1$ is defined by the standard immersion: $\phi \colon \underbrace \varphi _{\varphi \in R} \mapsto (\cos \varphi,\sin \varphi)$.
Let us notice that $\phi ^ * \Omega _1 (\varphi) = \frac{1}{{2\pi }}d\varphi = \frac{1}{{2\pi }}\eta$ where $\eta$ is the circle volume form.
Points $\varphi$ and $\varphi + 2m\pi$ where $m \in$ Z indeed define the same point on the circle.
Mapping of circle $S^1$ into circle $S^1$ we will get as the result of composition of mappings $\phi \circ f$, where $f$ is such smooth
function $f \colon R\to R$ so $f(\varphi + 2\pi) = f(\varphi) + 2n\pi$, where $n \in$ Z. Lets notice too that $\phi \circ f \colon \varphi \mapsto \phi
\left({f(\varphi)} \right) = \left({\cos f(\varphi),\sin f(\varphi)} \right)$ and
$\left({\phi \circ f} \right)^ * \Omega _1 = f^ * \circ \phi ^ * \Omega _1 = \frac{1}{{2\pi }}f^ * \eta = \frac{1}{{2\pi }}f'(\varphi)d\varphi$ so
the degree of mapping of a circle into a circle is given by the formula:
\begin{equation}
\deg f = \frac{1}{{2\pi }}\int\limits_0^{2\pi } {f'(\varphi)d\varphi = \frac{1}{{2\pi }}} [f(2\pi) - f(0)] = n.
\label{ref_eq05}
\end{equation}
For example if $f_n (\varphi) = n\varphi$; $n \in$ Z then obviously $\deg f_n = n$. It is to be easily seen that for such a mapping the image of the
circle $S^1$ $n$-times ''winds'' on the circles $S^1$.
That is the reason why number $n$ is called  the \textit{winding number}. Each mapping of a circle into a circle is homotopicaly equivalent to
some mapping $f_n$. 

So the set of all homotopy classes $\left\{ {S^1,S^1 } \right\}$ is bijective with the set of integer numbers, Z.
Analogously the set of all homotopy classes $\left\{ {S^n,S^n } \right\}$ is bijective with the set of integer numbers \textit{Z}.

Let us consider now $2$-dimensional sphere $S^2$.
The standard immersion $\phi$ is the mapping $\phi \colon \left({\theta,\varphi } \right) \mapsto
\left({\sin \theta \cos \varphi,\sin \theta \sin \varphi,\cos \theta } \right)$.

It is easy to see that $\phi ^ * \Omega _2 = \frac{1}{{4\pi }}\sin \theta d\theta \wedge d\varphi$, where $\sin \theta d\theta \wedge d\varphi$
is the form of the volume in sphere $S^2$. Mapping of the sphere $S^2$ into sphere $S^2$ is defined by the function
$f \colon \left({\theta,\varphi } \right) \mapsto \left({f_1 (\theta,\varphi),f_2 (\theta,\varphi)} \right)$.
This mapping is the result of the composition of mappings $\phi \circ f$:
 $\phi \circ f \colon \left({\theta,\varphi } \right) \mapsto \left({\sin f_1 (\theta,\varphi)
 \cos f_2 (\theta,\varphi),\sin f_1 (\theta,\varphi)\sin f_2 (\theta,\varphi),\cos f_1 (\theta,\varphi)} \right).$
After the simple transformation we get:
\begin{displaymath}
\left({\phi \circ f} \right)^ * \Omega _2 = f^ * \circ \phi ^ * \Omega _2 = \frac{1}{{4\pi }}f^ * \eta =
\frac{1}{{4\pi }}\sin f_1 (\theta,\varphi)\frac{{D\left({f_1,f_2 } \right)}}{{D\left({\theta,\varphi } \right)}}d\theta \wedge d\varphi.
\end{displaymath}
So the degree of the above mapping is given by the following formula:
\begin{equation}
\deg f = \frac{1}{{4\pi }}\int\limits_0^\pi {d\theta } \int\limits_0^{2\pi } {d\varphi \sin \left({f_1 (\theta,\varphi)} \right)}
\frac{{D\left({f_1,f_2 } \right)}}{{D(\theta,\varphi)}}d\theta d\varphi
\label{ref_eq06}
\end{equation}
Let an $n$-dimensional sphere $S^n$ be given in $R^{n + 1}$ by equation $(x^1)^2 +\cdots + (x^{n + 1})^2 = 1$. 
If $\hat f \colon S^n \to S^n$ is the smooth mapping from the sphere $S^n$ into sphere $S^n$ then the degree of this mapping has the form:
\begin{equation}
\deg \hat f = \frac{1}{{vol(S^n)}}\int\limits_{S^n } {\hat f^ * } \eta,
\label{ref_eq07}
\end{equation}
where $\eta$ is the volume form on $S^n$.
Let $\phi \colon R^n \supset D \to S^n$ be an immersion of sphere $S^n$ in $R^{n + 1}$ given like this:
 $\underbrace {\left({t^1,\ldots,t^n } \right)}_{ \in D} = t \mapsto \left({x^1 (t),\ldots,x^{n + 1} (t)} \right) \in S^n$.
Let us notice that $\Omega _n \left| {_{S^n } } \right. = \frac{1}{{vol(S^n)}}\eta$ where $\eta$ is the volume form on sphere $S^n$.

Instead of considering mapping $\hat f \colon S^n \to S^n$ we can consider its realization $f$ in parametrization $\phi$ which is defined by this
commutative diagram.
\begin{figure}[h]
\begin{center}
\includegraphics[width=2.5cm]{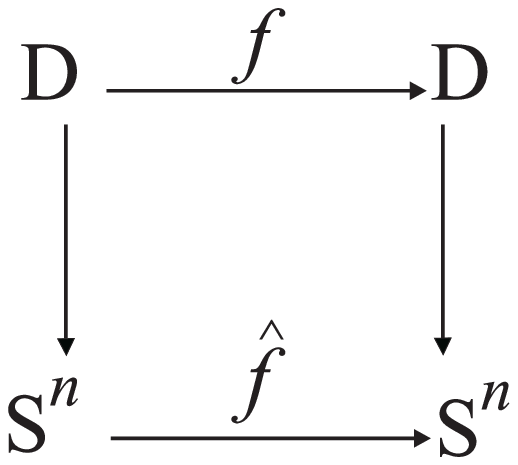}
\end{center}
\end{figure}


From this diagram it follows that $\hat f = \phi \circ f \circ \phi ^{ - 1}$ so
\begin{eqnarray*}
\deg \hat f & = &\frac{1}{{vol(S^n)}}\int\limits_{S^n } {\hat f^ * } \eta \nonumber = \int\limits_{S^n } {\hat f^ * } (\Omega _n)  \\
& = &\int\limits_{S^n } {\left({\phi \circ f \circ \phi ^{ - 1} } \right)} ^ * \left({\Omega _n } \right)  \\
& = &\int\limits_D {\phi ^ * \circ \phi ^{ - 1 * } } \circ f^ * \circ \phi ^ * (\Omega _n) \\
& = &\frac{1}{{vol(S^n)}}\int\limits_D {f^ * \eta (t)}
\end{eqnarray*}
where $\eta (t) = vol(S^n)\phi ^ * (\Omega _n)$ is the volume form on sphere $S^n$ in parametrization $\left({t^1,\ldots,t^n } \right)$.

The formula:
\begin{equation}
\deg f = \frac{1}{{vol(S^n)}}\int\limits_D {f^ * \eta (t)}
\label{ref_eq08}
\end{equation}
allows us to calculate the degree of mapping $\hat f \colon S^n \to S^n$ for some given parametrization of sphere.

\subsection{Solid angle form and index of vector field}

With any vector field defined in $R^{n + 1}$ we can connect its homotopic characteristics.
Let us consider the vector field $V(x) = V^\mu (x)\hat e{}_\mu$ $(\mu = 1,2,\ldots,n + 1)$
(where $\hat e_\mu$ is the orthonormal basis in $R^{n + 1}$) defined in the domain $U \subset$ $R^{n + 1}$.
Let us assume that this field takes the value zero (or is singular) at the interior points $p_1,p_2,\ldots,p_N$ belonging to $U$.
In this case in domain $U\backslash \left\{ {p_1,p_2,\ldots,p_N } \right\}$ we have the unit vector field $N(x) = \frac{{V(x)}}
{{\left| {V(x)} \right|}}.$

Let $M^n$ be any closed hypersurface lying in the domain $U$ and such that its interior contains the points $p_1,p_2,\ldots,p_N$.
The surface $M^n$ can be homotopic with sphere $S^n$.
Then, the Gauss mapping $\psi \colon M^n \to S^n$ is the mapping $\underbrace x_{ \in M^n } \mapsto N(x) \in S^n$.
The degree of this mapping is referred to as the index of the vector field $V$ in relation to hypersurface $M^n$ and is denoted by
$Ind_M V$ i.e. $Ind_M V = \deg \psi$.
We know that a form $\Omega _n$ restricted to sphere $S^n$, i.e. $\Omega _n \left| {_{S^n } } \right.$ is a form of a sphere $S^n$
volume shared by her volume $\Omega _n \left| {_{S^n } } \right. = \frac{1}{{vol(S^n)}}\eta$, where $\eta$ is a volume form of sphere $S^n$.
Because $\psi ^ * \left({\Omega \left| {_{S^n } } \right.} \right) = \frac{1}{{vol(S^n)}}\psi ^ * \eta$ so
\begin{displaymath}
\frac{1}{{vol(S^n)}}\int\limits_{M^n } {\psi ^ * } \eta = \frac{1}{{vol(S^n)}}\deg \psi \int\limits_{S^n } {\eta = \deg \psi }.
\end{displaymath}
Then, the homotopic characteristics of a vector field $V$, when this characteristics is its index in relation to given hipersurface $M^n$ is
given by the formula:
\begin{equation}
Ind_M V = \frac{1}{{vol(S^n)}}\int\limits_{M^n } {\psi ^ * } \eta = \int\limits_{M^n } {\psi ^ * \left({\Omega _n } \right)}
\label{ref_eq09}
\end{equation}
Let $\phi \colon R^n$ $\supset D \to R^{n + 1}$ be some parametrization of surface $M^n$ i.e. its immersion given as $\phi \colon \underbrace
{\left({t^1,t^2,\ldots,t^n } \right)}_{ \in D} = t \mapsto \left({x^1 (t),\ldots,x^{n + 1} (t)} \right) \in M^n$.
Then:
\begin{displaymath}\int\limits_{M^n } {\psi ^ * } \left({\Omega _n } \right) = \int\limits_D {\phi ^ * \circ \psi ^ * } \left({\Omega _n } \right) =
\int\limits_D {\left({\psi \circ \phi } \right)^ * \left({\Omega _n } \right)}.
\end{displaymath}
Since:
\begin{displaymath}
\underbrace {\left({t^1,\ldots,t^n } \right)}_{ \in D} = t \mapsto \left({N^1 (x(t)),\ldots,N^{n + 1} (x(t))} \right) \equiv
\left({N^1 (t),\ldots,N^{n + 1} (t)} \right).
\end{displaymath}
It follows that for the composition of mappings ${\psi \circ \phi }$ we have
\begin{displaymath}
\left({\psi \circ \phi } \right)^ * \left({\Omega _n } \right) = \frac{1}{{vol(S^n)}}\sum\limits_{\mu = 1}^{n + 1} {(- 1)^{\mu - 1} }
N^\mu (t)dN^1 \wedge\ldots \wedge d\mathop {N^\mu }\limits^ \vee \wedge\ldots \wedge dN^{n + 1}
\end{displaymath}
It is easy to see that:
\begin{displaymath}
\sum\limits_{\mu = 1}^{n + 1} {(- 1)^{\mu - 1} } N^\mu dN^1 \wedge\ldots \wedge d\mathop {N^\mu }\limits^ \vee \wedge\ldots
\wedge dN^{n + 1} = \frac{1}{{n!}} \in _{\mu _1 \mu _2\ldots \mu _{n + 1} } N^{\mu _1 } dN^{\mu _2 } \wedge\ldots \wedge dN^{\mu _{n + 1} }
\end{displaymath}
so
\begin{equation}
Ind_M V = \frac{1}{{n!vol(S^n)}}\int\limits_{M^n } { \in _{\mu _1 \mu _2\ldots \mu _{n + 1} } } N^{\mu _1 } dN^{\mu _2 }
\wedge\ldots \wedge dN^{\mu _{n + 1} }
\label{ref_eq10}
\end{equation}

For practical applications of this formula it is convenient to transform it into a form in which the components of the vector field  appear explicitly.
To that order let us notice that:
\begin{eqnarray*}
\frac{1}{n!} \in _{\mu _1 \mu _2\ldots\mu _{n + 1} } N^{\mu _1 } dN^{\mu _2 } \wedge\ldots \wedge dN^{\mu _{n + 1} } = \nonumber \\
\frac{1}{n!} \in _{\mu _1 \mu _2\ldots\mu _{n + 1} } N^{\mu _1 } \frac{{\partial N^{\mu _2 } }}{\partial t^{i_1 } }
\ldots\frac{{\partial N^{\mu _{n + 1} } }}{{\partial t^{i_n } }}dt^{i_1 } \wedge\ldots \wedge dt^{i_n } = \\
\in _{\mu _1 \mu _2\ldots\mu _{n + 1} } N^{\mu _1 } \frac{{\partial N^{\mu _2 } }}{{\partial t^1 }}\ldots\frac{{\partial N^{\mu _{n + 1} } }}
{{\partial t^n }}dt^1 \wedge\ldots \wedge dt^n.
\end{eqnarray*}
Since
\begin{displaymath}
\frac{{\partial N^\mu }}{{\partial t^i }} = \frac{\partial }{{\partial t^i }}\left[ {\frac{{V^\mu }}{{\left| V \right|}}} \right] =
\frac{1}{{\left| V \right|}}\frac{{\partial V^\mu }}{{\partial t^i }} - \frac{{V^\mu }}{{\left| V \right|^3 }}
\left[ {V^\alpha \frac{{\partial V^\alpha }}{{\partial t^i }}} \right],
\end{displaymath}
we have
\begin{displaymath}
(\psi \circ \phi)^ * \left({\Omega _n } \right) = \frac{1}{{vol(S^n)}} \in _{\mu _1 \mu _2\ldots\mu _{n + 1} }
\frac{{V^{\mu _1 } }}{{\left| V \right|^{n + 1} }}\frac{{\partial V^{\mu _2 } }}{{\partial t^1 }}\ldots
\frac{{\partial V^{\mu _{n + 1} } }}{{\partial t^n }}dt^1 \wedge\ldots \wedge dt^n.
\end{displaymath}
Hence, finally
\begin{equation}
Ind_M V = \frac{1}{{vol(S^n)}} \in _{\mu _1 \mu _2\ldots \mu _{n + 1} } \int\limits_D {\frac{{V^{\mu _1 } (t)}}
{{\left| V \right|^{n + 1} }}} \frac{{\partial V^{\mu _2 } }}{{\partial t^1 }}(t)\ldots \frac{{\partial V^{\mu _{n + 1} } }}
{{\partial t^n }}(t)dt^1\ldots dt^n.
\label{ref_eq11}
\end{equation}
We can also define index of the vector field $V$ at a point $p_i$,
where this field takes the value zero or is singular. To achieve
this we enclose the given point $p_i$ with a hypersurface $S_i$
that its interior contains exactly one point $p_i$. In such the
case $\int\limits_{S_i } {\psi ^ * } \left({\Omega _n } \right)$
is the integral formula for the index of vector field $V$ at the
point $p_i$. This index is denoted as $Ind_{p_i } V$. We will show
now that the index of the vector field $V$ in relation to the
hypersurface $M^n$ is equal to the sum of indices of this field at
the points $p_1,p_2,\ldots,p_N$, where this filed takes the value
zero (or is singular), and which lie inside the hypersurface
$M^n$. For this purpose, we take each point $p_i$ and enclose it
in a hypersurface $S_i$ which contains in its interior only the
point $p_i$. Let $\hat O$ denote one domain whose boundary is a
hypersurface $M^n$ and hypersurfaces $S_i$ $(i = 1,2..,N)$. Let 
$\hat \psi$ means the mapping $\hat \psi \colon \hat O \to S^n$ defined
as. The form $\Omega _n$ is closed hence $d[\hat \psi ^ *
\left({\Omega _n } \right)] = \hat \psi ^ * \left({d\Omega _n }
\right) = 0.$ Therefore using the Stokes theorem we have:
\begin{displaymath}
0 = \int\limits_{\hat O} {d[\hat \psi ^ * \left({\Omega _n } \right)] = \int\limits_{\partial \hat O} {\psi ^ * \left({\Omega _n } \right)} } =
\int\limits_{M^n } {\psi ^ * (\Omega _n }) - \sum\limits_{i = 1}^N {\int\limits_{S_i } {\psi ^ * \left({\Omega _n } \right)} }.
\end{displaymath}
From the last equation it follows that:
\begin{equation}
Ind_{M^n } V = \sum\limits_{i = 1}^N {Ind_{p_i } V}
\label{ref_eq12}
\end{equation}
From the above discussion we see that the index of vector field $V$ (defined on $R^n$) in point $x_0$ is given by a formula:
\begin{equation}
Ind_{x_0 } V = \frac{1}{{vol(S^{n - 1})}}\int\limits_S {\frac{1}{{\left| V \right|^n }}} \sum\limits_{\mu = 1}^n {(- 1)^{\mu - 1} }
V^\mu dV^1 \wedge\ldots \wedge d\mathop {V^\mu }\limits^ \vee \wedge\ldots \wedge dV^n,
\label{ref_eq13}
\end{equation}
where $S$ is a surface inside which lies the point $x_0$ at which a filed $V$ takes value zero (or is singular in it).
An example: when the vector field $V$ is defined in $R^2$ by a mapping $\left({x,y} \right) \mapsto \left({P(x,y),Q(x,y)} \right)$, so
\begin{equation}
Ind_{(x_0,y_0)} V = \frac{1}{{2\pi }}\oint\limits_C {\frac{{PdQ - QdP}}{{P^2 + Q^2 }}} = \frac{1}{{2\pi i}}\oint\limits_C {d\ln \left[
{\frac{f}{{\left| f \right|}}} \right]},
\label{ref_eq14}
\end{equation}
Let $f$ be a complex-valued function given by a formula $f(x + iy) = P(x,y) + iQ(x,y)$ and $C$ is a curve closed encircling point
$\left({x_0,y_0 } \right)$.

For example if a function $f(x + iy) = z^n$ $(n \in$ Z) i.e. when $f(z) = \left| {z^n } \right|e^{in\varphi }$ then
\begin{displaymath}
Ind_{\left({0,0} \right)} V = \frac{1}{{2\pi i}}\int\limits_0^{2\pi } {d\ln e^{in\varphi } } = \frac{n}{{2\pi }}\int\limits_0^{2\pi }
{d\varphi } = n.
\end{displaymath}
So the vector field $V$ defined on a plane by an analytic function $f(z) = P + iQ = \left({x + iy} \right)^n$ has at the origin the index
equal to n.
For instance, at the  origin, the vector field $V(x,y) = (x,y)$ has an index of 1, the vector field $V(x,y) = (x^2 - y^2,2xy)$ has an index of 2,
and the vector field $V(x,y) = \left({\frac{x}{{x^2 + y^2 }}, - \frac{y}{{x^2 + y^2 }}} \right)$ has an index of $- 1$.

\section{Homotopical classification of field equations solutions}

We are often able to make the (topological) homotopical
classification of solutions of a given system of differential
equations. With the solutions we associate the so called
typological index (topological quantum number), which we also call
the topological constant of motion. So defined index does not
change when we put the solution into the process of smooth (or
continuous) deformation i.e. the process of homotopy. Though in
the physical problems under consideration we need to ensure that
in the process of deformation the asymptotic of solution in
spatial infinity will not be changed. An example of very useful
topological index is a degree of mapping. The mapping is then a
solution of field equations. The homotopic index is characteristic
for a class of homotopically equivalent solutions. Solutions which
cannot be smoothly (continuously) deformed one into another have
different topological indices. The time evolution of a solution
coincident with the solution of equations of motion (field
equations)  can be treated as the smooth deformation of initial
conditions. If for time $t \to - \infty$ a solution is
characterized by some topological index so for time $t \to +
\infty$ this index will be not changed. Hence the topological
index is a constant of motion. However, it is different from the
constant of motion following from Noether's theorem since it has
nothing to do with the symmetries of the system.

\subsection{ 't Hooft-Polyakov monopole}

The 't Hooft-Polyakov monopole \cite{ref_Eguchi} is the static
spherically symmetric solution in the classical theory describing
Yang-Mills fields system and Higgs fields with gauge group
$SU(2)$. In this model the Higgs field takes  values in the Lie
algebra of the group $SU(2)\colon \phi = \phi ^a t^a$ where $a =
1,2,3$, $t^a =
 - \frac{i}{2}\sigma ^a$ ($\sigma ^a$ are Pauli matrices).
The following equations are fulfilled: $[t^a,t^b ] = \in ^{abc}
t^c$, $tr(t^a t^b) = - \frac{1}{2}\delta ^{ab}.$

On a classical level the system is defined by Lagrange function:
\begin{displaymath}
L = - tr[D_\mu \phi D_\nu \phi ]\eta ^{\mu \nu } + \frac{1}{2}tr\left({F_{\mu \nu } F_{\lambda \sigma } }
\right)\eta ^{\mu \lambda } \eta ^{\nu \sigma } - \frac{\lambda }{4}\left({\phi ^a \phi ^a - F^2 } \right)^2,
\end{displaymath}
where $D_\mu \phi = \partial _\mu \phi + g[A_\mu,\phi ]$, $A_\mu$ is the gauge field with values in the Lie algebra of $SU(2)$ ;
$A_\mu = A_\mu ^a t^a$, and $F_{\mu \nu } = \partial _\mu A_\nu - \partial _\nu A_\mu + g[A_\mu,A_v ]$
is a gauge field strength tensor.

Lagrange function does not change under gauge transformations:
\begin{eqnarray*}
A_\mu (x) & \mapsto & A_\mu ^\omega (x) = \omega A_\mu \omega ^{ - 1} - \frac{1}{g}\left({\partial _\mu \omega } \right)\omega ^{ - 1}, \nonumber \\
\phi (x) & \mapsto & \phi ^\omega (x) = \omega \phi \omega ^{ - 1},
\end{eqnarray*}
where $\omega = \omega (x) \in SU(2)$.
Euler-Lagrange equations have the following form:
\begin{eqnarray*}
D_\mu F^{\mu \nu } & = & g[D^\nu \phi,\phi ], \nonumber \\
D_\mu D^\mu \phi & = & - \lambda \left({\phi ^a \phi ^a - F^2 } \right)\phi,
\end{eqnarray*}
where $D_\mu F_{\lambda \sigma } = \partial _\mu F_{\lambda \sigma
} + g[A_\mu,F_{\lambda \sigma } ]$ and $\eta ^{\mu \nu } = diag(+
1, - 1, - 1, - 1)$ is Minkowski metric. For static solutions for
which $A_0 = 0$ the total energy of system is given by the
formula:
\begin{equation}
E = \int\limits_{R^3 } {d^3 x} \left\{ {\frac{1}{2}\left[ {B_i^a B_i^a + \left({D_i \phi } \right)^a \left({D_i \phi } \right)^a } \right] +
 \frac{\lambda }{4}\left({\phi ^a \phi ^a - F^2 } \right)^2 } \right\},
\label{ref_eq15}
\end{equation}
where $B_i = - \frac{1}{2} \in _{ijk} F_{jk}$ and $E_i = F_{0i}$. Conditions for the energy of system to be finite are such:
\begin{eqnarray}
A_i & \stackrel{\textstyle {\longrightarrow}}{\scriptscriptstyle {r\rightarrow\infty}} & f_i(\theta\ , \varphi)/r, \nonumber \\
\phi & \stackrel{\textstyle {\longrightarrow}}{\scriptscriptstyle {r\rightarrow\infty}} & F (\sin f_1(\theta\ , \varphi)\cos f_2(\theta\ , \varphi),
\nonumber \\
&& \sin f_1(\theta\ , \varphi) \sin f_2(\theta\ , \varphi), \cos f_1(\theta\ , \varphi)).
\label{ref_eq16}
\end{eqnarray}
It means that for $r \to \infty$ field $\phi$ is a mapping from a sphere $S_\infty ^2$ to sphere $S_\phi ^2$ with radius $F$.
The boundary conditions which guarantee the finiteness of the total energy $E$ define the  family of mappings from
$S_\infty ^2$ to $S_\phi ^2$ which are characterized by the index of the vector  field  $\left({\phi ^1,\phi ^2,\phi ^3 } \right)$ in
relation to sphere $S_\infty ^2$ i.e. (10):
\begin{equation}
Q = \frac{1}{{4\pi }}\frac{1}{2}\int\limits_{S_\infty ^2 } { \in _{abc} } \hat \phi ^a d\hat \phi ^b \wedge d\hat \phi ^c,
\label{ref_eq17}
\end{equation}
where:
\begin{displaymath}
\hat \phi ^a = \frac{{\phi ^a }}{{\left[ {(\phi ^1)^2 + (\phi ^2)^2 + (\phi ^3)^2 } \right]^{\frac{1}{2}} }}.
\end{displaymath}
The integer number $Q$ we call the topological quantum numbers.

\subsection{Instantons}

Euclidean Lagrange function for the $SU(2)$ pure gauge theory
\cite{ref_Eguchi} has the form:
\begin{displaymath}
L_E = - \frac{1}{{2g}}tr\left[ {F_{\alpha \beta } F_{\alpha \beta } } \right]
\end{displaymath}
(we sum up over repeating indices),
where
\begin{displaymath}
F_{\alpha \beta } = \partial _\alpha A_\beta - \partial _\beta A_\alpha + [A_\alpha,A_\beta ], \ \ \  A_\alpha = A_\alpha ^a t^a.
\end{displaymath}
This Lagrange function is invariant under a gauge transformation:
 $A_\alpha \mapsto A_\alpha ^\omega = \omega A_\alpha \omega ^{ - 1} - \left({\partial _\alpha \omega } \right)\omega ^{ - 1}$,
 where $\omega = \omega (x) \in SU(2)$.
Instantons are the solutions of Euler-Lagrange equations corresponding to the above Lagrange function for which Euclidean action
$S_E = \int\limits_{R^4 } {d^4 } xL_E$ is finite.
The sufficient condition for a solutions to have the finite action is:
\begin{displaymath}
A_a \stackrel{\textstyle {\longrightarrow}}{\scriptscriptstyle {|x|\rightarrow\infty}}\left(-\partial _a \omega \right) \omega ^{-1},
\end{displaymath}
where $\left| x \right| = \left[ {\left({x^1 } \right)^2 + \left({x^2 } \right)^2 + \left({x^3 } \right)^2 + \left({x^4} \right)^2 }
\right]^{\frac{1}{2}}$ and $\omega (x) \in SU(2)$.
Any element $\omega \in SU(2)$ can be written as
\begin{displaymath}
\omega = V^1 (i\sigma _1) + V^2 (i\sigma _2) + V^3 (i\sigma ^3) + V^4 1,
\end{displaymath}
where $\left({V^1 } \right)^2 + \left({V^2 } \right)^2 + \left({V^3 } \right)^2 + \left({V^4 } \right)^2 = 1$.

The asymptotic of gauge potential, $A_\alpha$, is defined by function $\omega$ which we can viewed as mapping from $S_\infty ^3$ to $SU(2)$.
The group $SU(2)$ as a manifold is the sphere $S^3$, so an element $\omega \in SU(2)$ for which $A_a \stackrel{\textstyle
{\longrightarrow}}{\scriptscriptstyle {|x|\rightarrow\infty}}\left(-\partial _a \omega \right) \omega ^{-1}$ defines the mapping
from $S_\infty ^3$ to $SU(2) \cong S^3$.
This, the set of gauge configurations which have finite, Euclidean action is defined by the boundary conditions which define the
mappings $S_\infty ^3 \to S^3$.
The set of such mappings falls into classes of homotopic mappings numbered by the degree of mentioned above mapping. The degree of this
mapping is equal to the index of vector field $\left({V^1,V^2,V^3,V^4 } \right)$ (defining element $\omega \in SU(2)$) in relation to
sphere $S_\infty ^3$. Hence the topological quantum number characteristic for the instanton solutions has the form of Eq.~(\ref{ref_eq10}):
\begin{equation}
Q = \frac{1}{{vol(S^3)3!}}\int\limits_{S_\infty ^3 } { \in _{\alpha \beta \gamma \delta } } V^\alpha dV^\beta \wedge dV^\gamma \wedge dV^\delta.
\label{ref_eq18}
\end{equation}
This number we can express by gauge field strength describing the instanton configuration. Making the simple transformation one can see that:
\begin{equation}
tr[d\omega \omega ^{ - 1} \wedge d\omega \omega ^{ - 1} \wedge d\omega \omega ^{ - 1} ] = - 2 \in _{\alpha \beta \gamma \delta } V^\alpha
dV^\beta \wedge dV^\gamma \wedge dV^\delta.
\label{ref_eq19}
\end{equation}
Since $vol(S^3) = 2\pi ^2$ it follows that $Q = - \frac{1}{{24\pi ^2 }}\int\limits_{S_\infty ^3 } {tr[d\omega \omega ^{ - 1} \wedge d\omega
\omega ^{ - 1} } \wedge d\omega \omega ^{ - 1} ]$.
Moreover: $tr\left({F \wedge F} \right) = dtr[A \wedge dA + \frac{2}{3}A \wedge A \wedge A]$,
where : $F = \frac{1}{2}F_{\mu \nu } dx^\mu \wedge dx^\nu$, $A = A_\mu dx^\mu$.

For instanton solutions, for $\left| x \right| \to \infty$ we have $F \to 0$ and because $F = dA + A \wedge A$, it follows that $\left| x
\right| \to \infty$ $dA \to - A \wedge A$ and $tr\left({F \wedge F} \right) \to - \frac{1}{3}dtr[A \wedge A \wedge A]$. Therefore for $\left| x
\right| \to \infty$ we have $A \to - d\omega \omega ^{ - 1}$ and $dtr[A \wedge A \wedge A] = - 3tr[F \wedge F]$.
Hence for $\left| x \right| \to \infty$ $dtr[d\omega \omega ^{ - 1} \wedge d\omega \omega ^{ - 1} \wedge d\omega \omega ^{ - 1} ] = 3tr[F \wedge F]$.
Now, we can easily show that:
\begin{equation}
Q = - \frac{1}{{8\pi ^2 }}\int\limits_{R^4 } {tr[F \wedge F]}.
\label{ref_eq20}
\end{equation}
Indeed
\begin{eqnarray*}
Q& = & - \frac{1}{24\pi ^2 } \int \limits_{S_\infty ^3 } {tr[d\omega \omega ^{- 1} \wedge d\omega \omega ^{- 1} d\omega \omega ^{ - 1} ] }  \\
& = & - \frac{1}{24\pi ^2 } \int\limits_{\partial R^4 } {tr[d\omega \omega ^{ - 1} \wedge d\omega \omega ^{ - 1} \wedge d\omega \omega ^{ - 1}  ]} \\
& = & - \frac{1}{24\pi ^2 }\int\limits_{R^4}
{ dtr [ d \omega \omega ^{ - 1} \wedge d\omega \omega ^{ - 1} \wedge d\omega \omega ^{ - 1} ]} \\
& = & - \frac{1}{8\pi ^2}\int\limits_{R^4 } {tr[F \wedge F]}.
\end{eqnarray*}
The last integral is called the Chern number.

Let us determine such element $\omega _{\left(k \right)} \in SU(2)$ that if  $A_a \stackrel{\textstyle {\longrightarrow}}{\scriptscriptstyle
{|x|\rightarrow\infty}} - \left(\partial \omega _{(k)} \right) \omega _{(k)}^{-1}$
then topological quantum number $Q$ is equal $k$ where $k \in$ Z.
In this purpose let us notice, that if $\omega = \omega _{\left(1 \right)} = \hat x^1 \left({i\sigma _1 } \right) + \hat x^2 \left({i\sigma _2 }
\right) + \hat x^3 \left({i\sigma _3 } \right) + \hat x^4$ 1, where $(\hat x^1)^2 +\ldots + (\hat x^4)^2 = 1$, then $Q = - \frac{1}{{24\pi ^2 }}
\int\limits_{S_\infty ^3 } {tr[d\omega _{\left(1 \right)} \omega _{(1)}^{ - 1} \wedge d\omega _{\left(1 \right)} } \omega _{\left(1 \right)}^{ - 1}
\wedge d\omega _{\left(1 \right)} \omega _{\left(1 \right)}^{ - 1} ] = 1$.
Moreover, we can show that:
\begin{eqnarray}
&tr&[d(\mathop \omega \limits_1 \mathop \omega \limits_2)(\mathop \omega \limits_1 \mathop \omega \limits_2)^{ - 1} \wedge d(\mathop \omega
\limits_1 \mathop \omega \limits_2)(\mathop \omega \limits_1 \mathop \omega \limits_2)^{ - 1} \wedge d(\mathop \omega \limits_1 \mathop \omega
\limits_2)(\mathop \omega \limits_1 \mathop \omega \limits_2)^{ - 1} ] \nonumber \\
=&tr&[d\mathop \omega \limits_1 \mathop \omega \limits_1 ^{ - 1} \wedge d\mathop \omega \limits_1 \mathop \omega \limits_1 ^{ - 1} \wedge d\mathop
\omega \limits_1 \mathop \omega \limits_1 ^{ - 1} ] \nonumber \\
+&tr&[d\mathop \omega \limits_2 \mathop \omega \limits_2 ^{ - 1} \wedge d\mathop \omega \limits_2 \mathop \omega \limits_2 ^{ - 1} \wedge d\mathop
\omega \limits_2 \mathop \omega \limits_2 ^{ - 1} ] + d\left\{ {(- 3)tr[(\mathop \omega \limits_1 \mathop \omega \limits_2)^{ - 1} d\mathop \omega
\limits_1 \wedge d\mathop {\omega ]}\limits_2 } \right\}.
\label{ref_eq21}
\end{eqnarray}
From this equation it follows that the degree of mapping $S_\infty ^3 \to SU(2)$
defined by element $\mathop \omega \limits_1 \mathop \omega \limits_2$ is equal to the sum of degrees of the mappings  defined by the elements
$\mathop \omega \limits_1$ and $\mathop \omega \limits_2$ because $\int\limits_{S_\infty ^3 } {} d\left\{ {(- 3)tr[(\mathop \omega \limits_1 \mathop
\omega \limits_2)^{ - 1} d\mathop \omega \limits_1 \wedge d\mathop {\omega ]}\limits_2 } \right\} = 0$.

Since the degree of mapping defined by a product $\mathop \omega \limits_1 \mathop \omega \limits_2$ is the sum of degrees defined by mappings
$\mathop \omega \limits_1$ and $\mathop \omega \limits_2$ the degree of the mapping  $[\omega _{\left(1 \right)} ]^k$ where $k \in$ Z is equal to $k$,
therefore $\omega _{\left(k \right)} = [\omega _{\left(1 \right)} ]^k$
i.e., vector field $V^\alpha$ in $R^4$ whose index in relation to surface $S_\infty ^3$ is equal $k$ is homotopic with vector field defined by a formula: $V^1 (i\sigma _1) + V^2 (i\sigma _2) + V^3 (i\sigma ^3) + V^4 1 = [\omega _{\left(1 \right)} ]^k$.

\section{Euler characteristics. Poincar\'{e}-Hopf theorem.}

Let $M$ be the $n$-dimensional Riemann manifold with a metric tensor
g \cite{ref_Straumann}. In a tangent bundle $TM$ we can locally
introduce the moving frame $\left\{ {e_a } \right\}$ $(a =
1,2,\ldots,n)$ such that $g\left({e_a,e{}_b} \right) \equiv
\left({e_a,e_b } \right) = \delta _{ab}$, moreover in the
contangent bundle $T^ * M$ we can define a moving cobasis $\left\{
{E^a } \right\}$ such that $E^a \left({e_b } \right) \equiv \left
\langle {E^a } \right.,\left. {e_b } \right\rangle = \delta _b^a$.
We define a local connection using the covariant derivative in
direction of the vector field $X$: $D_X e_a = e{}_b\omega _a^b
\left(X \right)$ where $\omega _a^b \left(X \right)$ is the value
of one form of local connection $\omega _a^b$ on a vector $X$. We
can extend the covariant derivative on any tensor field asking it
will fulfill the standard conditions. The exterior covariant
derivative is defined by: $De_a = e_b \otimes \omega _a^b$, where
$De_a$ is one form with vector values such that $\left\langle
{De_a } \right.,\left. X \right\rangle = D_X e{}_a$. We expect the
connection $\omega _a^b$ to be torsion free and metrical. Being
torsion free means that $dE^a + \omega _b^a \wedge E^b = 0$, and
the fact that the connection is metrical means that: $\omega _{ab}
+ \omega _{ba} = 0.$ In other words, the connection being metrical
means orthonormal moving frame its form is antisymmetric i.e. the
form of connection takes the values in Lie algebra $O(n)$ group. The
curvature two form for a given connection one form, in the
simplest way, we can define with the formula:
\begin{displaymath}
D^2 e_a = D[e_c \otimes \omega _a^c ] = \left({De_c } \right) \wedge \omega _a^c + e_c \otimes d\omega _a^c = e_b \otimes [d\omega _a^b + \omega _c^b
\wedge \omega _a^c ].
\end{displaymath}
The two-form $R_a^b = d\omega _a^b + \omega _c^b \wedge \omega _a^c$ is the local curvature two form corresponding to connection $\omega _a^b$.
From the connection being metrical we can easily see that it takes the value also in Lie algebra of group $O(n)$ i.e. $R_{ab} = - R{}_{ba}$.
When $n = 2$ it means that $R_{ab} = d\omega _{ab}$.
An exterior covariant derivative one can define for any p-form $\Omega$ taking the tensor values namely:
\begin{displaymath}
D\Omega _{b_1\ldots b_s }^{a_1\ldots a_k } = d\Omega _{b_1\ldots b_s }^{a_1\ldots a_k } + \sum\limits_{i = 1}^k {\omega _c^{a_i } } \wedge \Omega _
{b_1\ldots b_s }^{a_1\ldots c\ldots a_k } - (- 1)^p \sum\limits_{i = 1}^s {\Omega _{b_1\ldots c\ldots b_s }^{a{}_1\ldots a_k } }
\wedge \omega _{b_i }^c.
\end{displaymath}
It is easy to show that:
\begin{displaymath}
DR_b^a = dR_b^a + \omega _c^a \wedge R_b^c - R_c^a \wedge \omega _b^c = 0.
\end{displaymath}
The identity $DR_b^a = 0$ we call the Bianchi identity.

\subsection{The heuristic proof of the Gauss-Bonnet theorem
for the two-dimensional surfaces}

Let $M^2$ be a closed, oriented two dimensional manifold. We first
triangulate this surface i.e. we divide it into triangles in such
way that any two neighbouring triangles have one mutual
triangulation edge and more than 2 triangles meet at some vertex.
As any inside of a triangle, the so called the triangulation face,
has 3 edges, each one of them being the edge of the other face, so
for any triangulation the relationship $3F = 2E$ is true, where
$F$ means the number of faces, $E$ means the number of
triangulation edges, and $V$ is number of triangulation vertices.
To simplify further discussion the triangulation of a surface
$M^2$ we will choose in such a way that every triangulation edge
is a segment of some geodesics on the manifold $M^2$ treated as
Riemann manifold. If the triangulation is sufficiently dense
(exact), then each two points of the surface $M^2$ can be
connected with exactly one geodesics. Let us choose one of
triangulation triangles.

\begin{figure}
\begin{center}
\includegraphics[width=9.5cm]{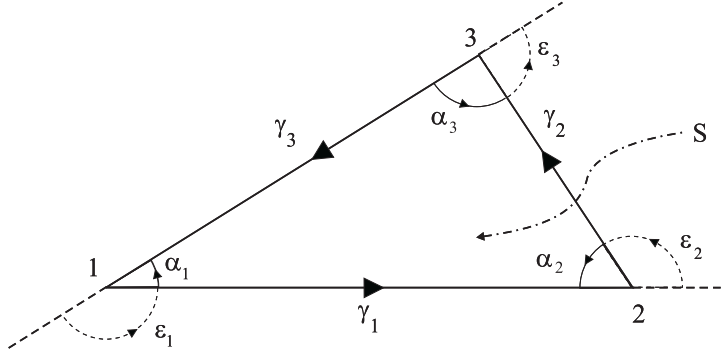}
\caption{\label{fig:Fig1}The way of counting angles in triangulation triangle.}
\end{center}
\end{figure}

We will prove that the sum of the internal angles $\alpha _1,\alpha _2,\alpha _3$ measured in a way shown in Fig.~1 fulfills the given Gauss relation:
\begin{equation}
\alpha _1 + \alpha _2 + \alpha _3 - \pi = \int\limits_S {R_{12} },
\label{ref_eq22}
\end{equation}
where $\gamma _1,\gamma _2,\gamma _3$ are the geodesics shown in Fig.~1, $R_{12}$ is the two-form of a curvature, $S$ is the face which edge (boundary)
is broken line $\gamma _1 \cup \gamma _2 \cup \gamma _3$.
Namely if $\gamma$ is geodesics then a vector field $X$ tangent to it fulfills the equation $D_X X = 0$.
At every point of a geodesics one can introduce the orthonormal moving frame $e_1,e_2$ and define an angle $\theta$ between vectors $e_1$ and $X$ what
is shown in Fig.~2.

\begin{figure}
\begin{center}
\includegraphics[width=9.5cm]{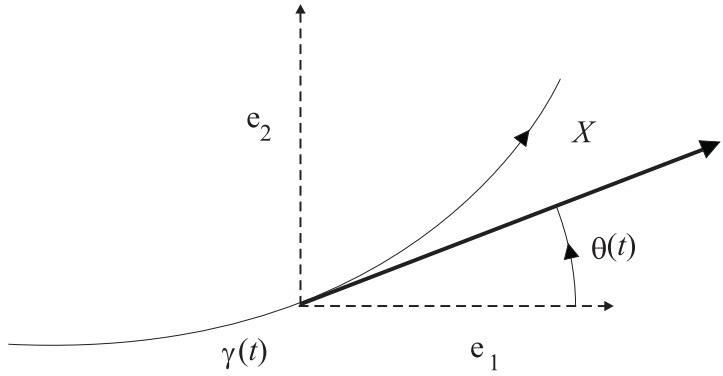}
\caption{\label{fig:Fig2}Definition of an angle $\theta$ (t is a parameter on a geodesics).}
\end{center}
\end{figure}

Then $X(t) = e_1 \cos \theta (t) + e_2 \sin \theta (t)$.
Because $D_X X = 0$ so
\begin{displaymath}
e_1 \sin \theta [\omega _{12} (X) - \frac{{d\theta }}{{dt}}] + e_2 \cos \theta [\frac{{d\theta }}{{dt}} - \omega _{12} (X)] = 0.
\end{displaymath}
i.e. $\frac{{d\theta }}{{dt}}(t) = \omega _{12} (X)$ where $\omega _{12} (X) = \left({e_1,D_X e_2 } \right)$ and $\omega _{12} = \left({e_1,De_2 }
\right)$
is connection one form on manifold $M^2$.
To get the Gauss relation Let us consider an integral
\begin{eqnarray*}
\oint\limits_{\gamma _1 \cup \gamma _2 \cup \gamma _3 } {d\theta}
&=& [\theta _2 (in) - \theta _1 (out)] + [\theta _3 (in) - \theta _2 (out)] + [\theta _1 (in) - \theta _3 (out)] \\
&=& [\theta _1 (in) - \theta _1 (out)] + [\theta _2 (in) - \theta _2 (out)] + [\theta _3 (in) - \theta _3 (out)] \\
&=& \psi _1 + \psi _2 + \psi _3,
\end{eqnarray*}
where $\psi _i = \theta _i (in) - \theta _i (out)$ for $i = 1,2,3.$

\begin{figure}
\begin{center}
\includegraphics[width=9.5cm]{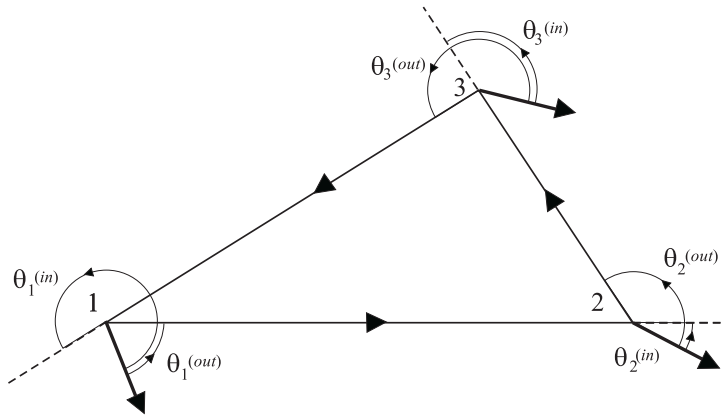}
\caption{\label{fig:Fig3}Way of counting angles in  the geodesics triangle.}
\end{center}
\end{figure}

From Fig. 3 and Fig. 1 we can see that: $\psi _1 = 2\pi - \in _1 $,$\psi _2 = - \in _2, $ $\psi _3 = - \in _3 $ and $\in _i + \alpha _i = \pi$ for
$i = 1,2,3$ - so $\psi _1 = \pi + \alpha _1 $, $\psi _2 = \alpha _2 - \pi$, $\psi _3 = \alpha _3 - \pi$ and hence:
\begin{displaymath}
\oint\limits_{\gamma _1 \cup \gamma _2 \cup \gamma _3 } {d\theta = \alpha _1 } + \alpha _2 + \alpha _3 - \pi = \oint\limits_{\gamma _1 \cup \gamma _2
\cup \gamma _3 } {\omega _{12} } = \int\limits_S {d\omega _{12} } = \int\limits_S {R_{12} }.
\end{displaymath}
Triangulation defines the sequence of the geodesics triangles $S_1,S_2,\ldots,S_F$,
where $F$ is a number of faces.
If with $(\alpha _1^i,\alpha _2^i,\alpha _3^i)$ we will denote the interior angles in i - th geodesics triangle then
\begin{displaymath}
\sum\limits_{i = 1}^F {\left({\alpha _1^i + \alpha _2^i + \alpha _3^i - \pi } \right)} = \sum\limits_{i = 1}^F {\int\limits_{S_i } {R_{12} } } =
\int\limits_{M^2 } {R_{12} } = 2\pi V - \pi F.
\end{displaymath}
From relation $3F = 2E$ we have $F = 2\left({E - F} \right)$ so
$\int\limits_{M^2 } {R{}_{12}} = 2\pi (F - E + V)$ or equivalently:
\begin{displaymath}
F - E + V \equiv \chi \left({M^2 } \right) = \frac{1}{{2\pi }}\int\limits_{M^2 } {R_{12} }.
\end{displaymath}
The integer number $\chi (M^2)$ is called the Euler characteristics of manifold $M^2$.
The determined relation between the Euler characteristics of manifold $M^2$ and an integral of curvature two-form $R_{12}$ over this manifold we
call the Gauss-Bonnet theorem.
The Euler characteristic of manifolds is an intrinsic property of manifolds, and does not depend either on how we choose to triangulate $M^2$, or on
how we choose the connection forms $\omega _{12}$. Moreover it can bee shown that two compact and closed surfaces $M_1$, $M_2$ are homeomorphic iff
these surfaces are orientable (or not) and when their Euler characteristics are equal. This is a very important theorem about topological classification
of the two dimensional surfaces.
 Any compact orientable closed surface is homeomorphic with connected sum of some number g of the tori $T^2$ and a sphere $S^2$ :
 $M^2 \cong S^2 \#T^2\#\ldots\#T^2$
We say that such a surface that it has a genus g and then its Euler characteristics is $2 - 2g$.
The connected sum $M_1 M_2$ of manifold $M_1$ and $M_2$ we construct as follows:
a) we cut out the little ball in each manifold,
b) we glue this manifolds along the edges of the mentioned above balls.
Its easy to see that $\chi (M_1 \# M_2)= \chi (M_1) + \chi (M_2) - 2$.

\subsection{Vector field index on manifolds and Poincar\'{e} theorem}

\textit{Theorem:}
Let $M^2$ be a closed, compact and orientable manifold:, let $V$ be the smooth vector field on $M^2$, let $p_1,\ldots p_R$ be the isolated
points at which the field $V$ takes zero value and $Ind_{p_i } V$ be the mean mean index of the vector field $V$ at point $p_i$. Then
\begin{displaymath}
\sum\limits_{i = 1}^R {Ind_{p_i } } V = \frac{1}{{2\pi }}\int\limits_{M^2 } {R_{12} } = \chi (M^2).
\end{displaymath}
\textit{Proof:}\\
Let $p_i$ be one of the points where the vector field $V$ takes value zero. In the neighborhood of this point exists the orthonormal moving
frame $e_1,e_2$. Besides everywhere (locally) outside the points $p_1,p_2,\ldots,p_R$ we can define another orthonormal moving frame
$\hat e_1,\hat e_2$ such that $\hat e_1 = \frac{V}{{\left| V \right|}}$ and $\hat e_2$ is the unit vector orthogonal to $\hat e_1$.
When at every point of manifold $M^2$ in which both moving frame are defined, the matrix of passage from one to the other is an element of
orthogonal group with the determinant equal with $+1$. So
\begin{eqnarray*}
\left({e_1,e_2 } \right) = \left({\hat e_1,\hat e_2 } \right)\left[ {\begin{array}{*{20}c}
{\cos \alpha } & { - \sin \alpha } \\
{\sin \alpha } & {\cos \alpha } \\
\end{array}} \right].
\end{eqnarray*}
The forms of the metric connection with respect to the moving frames $(e_1,e_2)$ and $\left({\hat e_1,\hat e_2 } \right)$ we define as:
$De_1 = e_2 \otimes \omega _{21}$, $D\hat e_1 = \hat e_2 \otimes \hat \omega _{21}$.
From the formula (10) results that:
\begin{equation}
Ind_{p_i } V = \frac{1}{{2\pi }}\oint\limits_C {[(\hat e_1,e_1 })d(\hat e_1,e_2) - (\hat e_1,e_2)d(\hat e_1,e_1)],
\label{ref_eq23}
\end{equation}
where $C$ is a curve closed encircling point $p_i$.
We can calculate directly that:
\begin{equation}
(\hat e_1,e_1)d(\hat e_1,e_2) - (\hat e_1,e_2)d(\hat e_1,e_1) = d\alpha = \omega _{12} - \hat \omega _{12}.
\label{ref_eq24}
\end{equation}
Because $dd\alpha = 0$ so $d\omega _{12} = d\hat \omega _{12} = R_{12}$.
Hence:
\begin{displaymath}
Ind_{p_i } V = \frac{1}{{2\pi }}\oint\limits_C {d\alpha = \frac{1}{{2\pi }}} \oint\limits_C {[\omega _{12} } - \hat \omega _{12} ].
\end{displaymath}
To determine the sum of indices of the vector field $V$ connected with all points in which this fields takes the zero value, we must show that the
surface $M^2$ is of the form $M^2 = M_ + \cup M_ - $, where $M_ + $ is a subset in $M^2 $ compounded of the neighborhoods of this points where the
field $V$ takes the value of zero
(each such neighborhood contains exactly one zero of the field $V$), $M_ - $ is the complement of a set $M_ + $ in $M^2$.
Hence:
\begin{eqnarray*}
\sum\limits_{i = 1}^R {Ind_{p_i } } V &=& \frac{1}{{2\pi }}\int\limits_{\partial M_ + } {[\omega _{12} } - \hat \omega _{12} ] \\
&=& \frac{1}{{2\pi }}\int\limits_{\partial M_ + } {\omega _{12} } + \frac{1}{{2\pi }}\int\limits_{\partial M_ - } {\hat \omega _{12} } =
\frac{1}{{2\pi }}\left[\int\limits_{M_ + } {d\omega _{12} } + \int\limits_{M_ - } {d\hat \omega _{12} } \right] \\
&=& \frac{1}{{2\pi }}\int\limits_{M^2 } {R_{12} } = \chi (M^2).
\end{eqnarray*}
We can see then that the sum of indices of any smooth vector field defined on a manifold $M^2$ does not depend from the choice of this field.
This is the topological characteristics of the manifold $M^2$.

The Hopf-Poincar\'{e} theorem is the truth for $n=2r$ dimensional closed and compact manifolds though its proof in the general cases is much more
difficult than for the two-dimensional case.
If $V$ is the smooth vector field on $n$-dimensional Riemann manifold $M^n$ with takes the value zero in an isolated point $p_0 \in M^n$ so we
can surround this point with some $(n-1)$-dimensional closed manifold $S$ to the inside of which belongs a point $p_0 $.
If $(\hat e_1,\hat e_2,\ldots,\hat e_n)$ is the orthogonal moving frame defined in such neighborhood of a point $p_0$ which
contains the submanifold $S$, then the index of the vector field $V$ in a point $p_0$ is given by a formula:
\begin{displaymath}
Ind_{p_0 } V = \frac{1}{{vol(S^{n - 1})(n - 1)!}}\int\limits_S { \in ^{a_1 a_2\ldots a_n } } (e_1,\hat e_{a_1 })d(e_1,\hat e_{a_2 })
\wedge\ldots \wedge d(e_1,\hat e_{a_n }),
\end{displaymath}
where $e_1 = \frac{V}{{\sqrt {(V,V)} }}$.

The above definition does not depend either on how we choose the submanifold $S$ or on how we choose a moving frame
$\left({\hat e_1,\hat e_2,\ldots,\hat e_n } \right)$.
If a vector field $V$ vanishes at points $p_1,p_2,\ldots.p_N \in M^{2r}$ then we can show that:
\begin{equation}
\sum\limits_{i = 1}^N {Ind_{p_i } } V = \int\limits_{M^{2r} } {e(R)}
\end{equation}
where $\int\limits_{M^{2r} } {e(R)}$ is the Euler characteristics of a manifold $M^{2r}$ when
 $e(R) = \frac{1}{{(4\pi)^r r!}}\eta ^{a_1 b_1\ldots a_r b_r } R_{a_1 b_1 } \wedge\ldots \wedge R_{a_r b_r }$ is the Euler
 form of the manifold $M^{2r}$,
 $R_{ab} = \frac{1}{2}R_{ab\mu \nu } dx^\mu \wedge dx^\nu$ is curvature two-form and $\eta ^{a_1 b_1\ldots a_r b_r } =
 \frac{1}{{\sqrt g }} \in ^{a_1 b_1\ldots a_r b_r }$.
The proof of his theorem is analogous to the proof of the Hopf-Poincar\'{e} theorem in two dimensions, and is left as a
(tedious) exercise to the reader.
So:
\begin{enumerate}
    \item[a)] An integral $\int\limits_{M^{2r} } {e(R)}$ does not depend from a choice of a metric $g$ and is equal to an integral number.
    \item[b)] A sum $\sum\limits_{i = 1}^N {Ind_{p_i } } V$ takes the same value for any smooth vector field on $M^{2r}$ having a finite number
    of isolated zeros.
    \item[c)] If an integral $\int\limits_{M^{2r} } {e(R)}$ is different from zero then $M^{2r}$ does not have a smooth vector field not having
    the zero points.
\end{enumerate}
An integral $\int\limits_{M^{2r} } {e(R)}$ we call Euler characteristic of manifold $M^{2r}$ and we mark it with the symbol $\chi (M^{2r})$.
For the compact manifold the vanishing of Euler characteristics is the sufficient condition for existence pseudoriemanian metric of signature 1.
Let the manifold $M^n$ be equipped with Riemann metric (,). According to the assumption there exists on $M^n$ the smooth unit vectors field $n$.
On $M^n$ we define the quadratic form:
\begin{displaymath}
\left({V - n(V,n)} \right)^2 - \left({V,n} \right)^2
\end{displaymath}
where $V$ is any vector field on $M^n$ , and $V^2 = (V,V)$.
Defined in this way quadratic form has a signature 1.

Euler characteristics can be interpreted as an obstruction for constructing a smooth field on $M^{2r}$ which nowhere takes the value of zero.
Generalization of this observation led to the creation of so called theory of the characteristic classes.

\section{Conclusions}

In this paper we provided a pedagogical discussion of topological
quantum numbers from the perspective of the degree of a smooth
mapping between smooth manifolds. The effective use of
differential forms allowed us to give a novel derivation of the
index of a vector field.

This in turn allowed us to derive the topological quantum number
characterizing e.g. monopole configurations in the Yang-Mills-Higgs 
theory with gauge group $SU(2)$. We also demonstrated
the bijectivity between the elements of the $SU(2)$ gauge group
and unit vector fields on $R^4$.

Then we presented an original, elementary proof of Gauss-Bonnet
and Poincar\'{e}-Hopf theorems for compact, closed, oriented, 
two dimensional manifolds. Unlike in the traditional proofs we 
used effective language of differential forms and
skilfully used the notion of the form of connection. The new
ingredient in the proof of the Poincar\'{e}-Hopf theorem was the
expression of an index of the vector field at a point by suitable
scalar products. This line of proof (after a generalization to
higher dimensions) creates a possibility of formulation a
sufficient condition for the existence of a pseudo-Riemannian
metric on compact manifolds without a boundary. Such a condition
would be very useful in modern cosmology. Finally we discuss the
Poincar\'{e}-Hopf theorem for any even dimensional compact
orientable manifold without boundary.

\subsection*{Acknowledgements}
M.S. acknowledges the support by the Marie Curie Actions Transfer of Knowledge
project COCOS (contract MTKD-CT-2004-517186).

\end{document}